\documentclass[preprint,preprintnumbers,amsmath,amssymb]{revtex4}
\usepackage{amssymb,latexsym,amsmath}

\usepackage[activeacute,english]{babel}
\usepackage{fancyhdr}
\usepackage{textcomp}
\usepackage{amsfonts}
\usepackage{graphicx}
\usepackage{subfigure}
\usepackage{graphics}
\usepackage{color}
\usepackage{array}

\usepackage[active]{srcltx}

\newcommand{\al}{\alpha}
\newcommand{\pa}{\partial}
\newcommand{\ep}{\epsilon}

\newcommand{\De}{\Delta}
\newcommand{\vphi}{\varphi}

\newcommand{\rar}{\rightarrow}

\newcommand{\non}{\nonumber}

\begin{document}

\title{Two charges on plane in a magnetic field: II. Moving neutral quantum system across a magnetic field}

\author{M.A.~Escobar-Ruiz}
\email{mauricio.escobar@nucleares.unam.mx}
\author{A.V.~Turbiner}
\email{turbiner@nucleares.unam.mx}
\affiliation{Instituto de Ciencias Nucleares, Universidad Nacional
Aut\'onoma de M\'exico, Apartado Postal 70-543, 04510 M\'exico,
D.F., Mexico}

\date{June 27, 2014}

\begin{abstract}
The moving neutral system of two Coulomb charges on a plane subject to a constant magnetic field $B$ perpendicular to the plane is considered. It is shown that the composite system of finite total mass is bound for any center-of-mass momentum $P$ and magnetic field strength; the energy of the ground state is calculated accurately using a variational approach. Their accuracy is cross-checked in a Lagrange-mesh method for $B=1$\,a.u. and in a perturbation theory at small $B$ and $P$. The constructed trial function has the property of being a uniform approximation of the exact eigenfunction. For a Hydrogen atom and a Positronium a double perturbation theory in $B$ and $P$ is developed and the first corrections are found algebraically.
A phenomenon of a sharp change of energy behavior for a certain center-of-mass momentum and a fixed magnetic field is indicated.

\end{abstract}


\maketitle

\begin{center}
\section*{Introduction}
\end{center}

\hskip 1cm
Two-dimensional planar systems of several Coulomb charges of finite masses, both classical and quantum, exhibit many interesting properties which are usually either hidden or absent in three dimensional one. The situation gets even more interesting if a magnetic field perpendicular to the plane is imposed. The first non-trivial case of two Coulomb charges reveals the existence of unexpected concentric classical trajectories and straight line motion (for neutral systems) (see \cite{ET:2013} and references therein) in classical problem. In the quantum case there exist a certain eigenstates which are characterized by extra quantum numbers. They are the eigenfunctions of the operators which are not global integrals \cite{ET-q:2013}. Usually, these eigenfunctions are of polynomial form (up to a multiplicative factor) \cite{Taut1,Taut2,ET-q:2013}. In two important cases of quasi-equal charges $(\frac{e_1}{m_1}=\frac{e_2}{m_2})$ and neutral system at rest the hidden algebra $sl(2)$ in finite-dimensional representation is present for a certain discrete values of a magnetic field strength \cite{Turbiner:1994,ET-q:2013}. It is needless to say that in the three-dimensional case there exists the phenomenon of dissociation which is absent in the two-dimensional planar case. All that indicates
the absence of a trivial connection between two- and three-dimensional cases. However, it turns out that there exists a number of properties which are similar for two- and three-dimensional systems in a constant uniform magnetic field. This similarity will be mentioned in our presentation.

\hskip 1cm
In our previous paper \cite{ET-gen:2013} an accurate variational solution for several low-lying states for both quasi-equal charges $(\frac{e_1}{m_1}=\frac{e_2}{m_2})$ and neutral system at rest was given. The stability of the system was shown for all magnetic fields.
The accuracy of the obtained results was evaluated in a specially designed, convergent perturbation theory. The goal of the present paper is to perform the study of the ground state of a moving neutral system of two Coulomb charges of finite masses in a wide range of magnetic fields and to check its stability. It has to be emphasized that the global integral of the motion, the Pseudomomentum ${\bf K}$ established by Gor'kov \& Dzyaloshinskii \cite{GD:1967} for the three-dimensional case remains the global integral for planar case as well, it coincides with the center-of-mass (CMS) canonical momentum of the system, ${\bf P} = {\bf K}$. We focus on a particular case of the Hydrogen atom (the case of unequal masses) and we also mention Positronium (the case of equal masses). To the best knowledge of the present authors there was a single attempt so far to study the moving Hydrogen atom on the plane carried out by Lozovik \textit{et al}, \cite{Lozovik:2002}. It was done for very weak magnetic fields. This problem is seen as an important for solid state physics, for the physics of excitons.

\hskip 1cm
As a first step we consider the case of weak magnetic fields and small CMS momentum in double perturbation theory in powers of $B$ and $P$. It will be realized in the "nonlinearization procedure" \cite{Turbiner:1979-84}, in the framework of which the computation of the coefficients of the perturbation series is a purely algebraic problem. In the case of larger fields and CMS momenta we are going to employ the variational formalism constructing the trial function in such a way to combine into an interpolation a WKB expansion at large distances with perturbation theory expansion at small distances near the extremum of the potential \cite{Turbiner:1988-2010}.

\bigskip

\section{Generalities}

The Hamiltonian, which describes a two-body neutral system, $(e, m_1), (-e, m_2)$ assuming $e>0$, in a constant and uniform magnetic field $B$ perpendicular to the plane, has of the form,
\begin{equation}
\begin{aligned}
{ {\hat H}} = &  \frac{{({\mathbf {\hat p}_1}-e\,{\mathbf A_1})}^2}{2\,m_1} + \frac{{({\mathbf {\hat p}_2}+e\,{\mathbf A_2})}^2}{2\,m_2}
 - \frac{e^2}{\mid {\boldsymbol \rho}_1 - {\boldsymbol \rho}_2 \mid} \,,\qquad \boldsymbol{\rho}_{1,2} \in \Re^2\ ,
\label{Hcar}
\end{aligned}
\end{equation}
where $\hslash=c=\frac{1}{4\,\pi\, \ep_0}=1$, ${\mathbf {\hat p}}_{1,2}=-i\,\nabla_{1,2}$ is the
momentum and ${\boldsymbol \rho}_{1,2}$ is the position vector of the first (second) particle.
Here the symmetric gauge $\mathbf A_{1,2}=\frac{1}{2}\,\mathbf B\times {\boldsymbol \rho}_{1,2}$
is chosen. It is easy to check that the total Pseudomomentum,
\begin{equation}
 {\mathbf {\hat K}}\, =\, \mathbf {\hat p}_1  + e\,\mathbf A_{1} +
 \mathbf {\hat p}_2 - e\,\mathbf A_{2}\ ,
\label{pseudo}
\end{equation}
is a vector integral of motion belonging the plane, on where the dynamics is developed,
\[
      [ {\hat H}\ ,\ {\mathbf {\hat K}}]\ =\ 0\ ,
\]
as well as the total angular momentum
\begin{equation}
\boldsymbol  {\hat L} = {\boldsymbol \rho}_1 \times {\mathbf {\hat p}}_1+ {\boldsymbol \rho}_2\times {\mathbf {\hat p}}_2\ ,
\label{Lz}
\end{equation}
$[\, \boldsymbol  {\hat L}, \,  {\hat H} \,]=0$. The vector $\boldsymbol  {\hat L}$ is perpendicular to the plane.
In general, the problem is not completely integrable.

It is convenient to introduce center-of-mass (c.m.s.) coordinates
\begin{equation}
\begin{aligned}
&\mathbf R = \mu_1\, {\boldsymbol \rho}_1 + \mu_2\,{\boldsymbol \rho}_2 \ ,
\quad  {\boldsymbol \rho}= {\boldsymbol \rho}_1 - {\boldsymbol \rho}_2\ ,
\\ & \mathbf {\hat P} = {\mathbf {\hat p}}_1 + {\mathbf {\hat p}}_2 \ ,
\qquad \quad \, \, {\mathbf {\hat p}} = \mu_2\,{\mathbf {\hat p}}_1 -  \mu_1\,{\mathbf {\hat p}}_2\ ,
\end{aligned}
\label{CMvar}
\end{equation}
where $M = m_1 + m_2$ is the total mass of the system and $\mu_i=\frac{m_i}{M}$ is relative mass of the $i$th charge. In these coordinates
\begin{equation}
\mathbf {\hat K}  = \mathbf {\hat P}\ +\ e\,\mathbf A_{{\boldsymbol \rho}}  \ ,
\label{pseudoR}
\end{equation}
\begin{equation}
\boldsymbol {\hat L}\ =\ ({\mathbf R} \times {\mathbf {\hat P}}) + ({\boldsymbol \rho}\times \mathbf {\hat p})
\equiv \mathbf {\cal{\hat L}} + \boldsymbol {\hat \ell} \ ,
\label{LzR}
\end{equation}
(cf. (\ref{pseudo}), (\ref{Lz})), where $\mathbf A_{{\boldsymbol \rho}}= \frac{1}{2} \mathbf B\times {\boldsymbol \rho}$.

It is easy to check that the integrals $\mathbf {\hat K}=({\hat K}_x, {\hat K}_y),\ \boldsymbol {\hat L} = {\hat L}\ \mathbf {n_z}$ obey the commutation relations
\begin{equation}
\begin{aligned}
&[ {\hat K}_x,\,{\hat K}_y ] = 0 \,,
\\ & [ {\hat L} ,\,{\hat K}_x ] = {\hat K}_y \,,
\\ & [ {\hat L},\,{\hat K}_y ] = -{\hat K}_x \,,
\label{AlgebraInt}
\end{aligned}
\end{equation}
hence, they span a noncommutative algebra with the Casimir operator ${\cal {\hat C}}$,
\begin{equation}
{\cal {\hat C}}\ =\ {\hat K}_x^2+{\hat K}_y^2 \ .
\label{Casimir}
\end{equation}

It is convenient to make unitary transformation of the canonical momenta
\[
  U^{-1}\,{\mathbf {\hat P}}\, U \ = \ {\mathbf {\hat P}} - e\,\mathbf A_{\boldsymbol \rho}  \quad , \quad
U^{-1}\,{\mathbf {\hat p}}\, U \ = \ {\mathbf {\hat p}} + e\,\mathbf A_{\mathbf R}
\ ,
\]
with
\begin{equation}
\label{U}
 {U}\ =\ e^{-i\,e\,\mathbf A_{\boldsymbol \rho}\cdot \mathbf R} \ .
\end{equation}
Then, the unitary transformed Pseudomomentum reads
\begin{equation}
  {\mathbf K^{\prime}}\ =\ U^{-1}\,{\mathbf {\hat K}} \, U \ = \  {\mathbf {\hat P}} \ ,
\label{KTrans}
\end{equation}
it coincides with the c.m.s. momentum of the whole, composite system, see (\ref{pseudo}).
The unitary transformed Hamiltonian (\ref{Hcar}) takes the form
\begin{equation}
  {\cal {\hat H}}^{\prime} \ =\  U^{-1}\,{\cal {\hat H}}\, U \ = \ \frac{ {( \mathbf {\hat P} - e\,\mathbf B\times {\boldsymbol \rho})}^2}{2\,M}
  +\frac{{({\mathbf {\hat p}}- {q_\text{w}}\,{\mathbf A_{\boldsymbol \rho}})}^2}{2\,m_{r}} - \frac{e^2}{\rho}\ ,
\label{H}
\end{equation}
here $q_{\rm{w}} = e\,(\mu_2^2 - \,\mu_1^2)=e (\mu_2 - \,\mu_1)$  is an effective charge (weighted total charge).
It is evident, $[\, \mathbf {\hat K}^{\prime}, \, {\cal {\hat H}}^{\prime} \,]=0$. The eigenfunctions of  ${\cal {\hat H}}^{\prime}$ and ${\cal {\hat H}}$ are related
\begin{equation}
   \Psi^{\prime}\ =\ \Psi\ e^{ i\,e\,\mathbf A_{\boldsymbol \rho}\cdot \mathbf R}\ .
\label{psiprime}
\end{equation}

It is easy to check that the eigenfunction of $\mathbf {\hat P}$ has the form
\begin{equation}
 \Psi^{\prime}_{{}_{\mathbf P}}(\mathbf R\,, \boldsymbol \rho)\ = \ \text{e}^{i\,\mathbf P\cdot \mathbf R}\,\psi_{{}_{\mathbf P}}(\boldsymbol \rho)    \ ,
\label{psik}
\end{equation}
where $\bf P$ is the eigenvalue and $\psi_{{}_{\bf P}}(\boldsymbol \rho)$ depends on the relative coordinate $\boldsymbol \rho$.
Substituting $\Psi^{\prime}_{{}_{\mathbf P}}$ into the Schr\"odinger equation for ${\cal {\hat H}}^{\prime}$ we obtain the equation of the relative motion
\begin{equation}
\label{HKa}
 h\,\psi_{\mathbf P}(\boldsymbol \rho)\ \equiv \
 \bigg[
 \frac{{({\mathbf {\hat p}}-e\,(\mu_2-\mu_1)\,{\mathbf A_{\boldsymbol \rho}})}^2}{2\,m_{r}} +
 \frac{ {( \mathbf {P}-\,e\,\mathbf B\times {\boldsymbol \rho} )}^2}{2\,M}   -\frac{e^2}{\rho}
 \bigg]\
 \psi_{\mathbf P}(\boldsymbol \rho)\ = E\ \psi_{\mathbf P}(\boldsymbol \rho)\ ,
\end{equation}
where CMS momentum $\bf P$ plays a role of external parameter.

The equation (\ref{HKa}) is the basic equation we are going to study. At $\bf P=0$, the relative angular momentum $\hat \ell_z= - i \,\pa_{\varphi}$ is conserved, the relative polar angle $\varphi$ is separated out and the problem (\ref{HKa}) is reduced to a study of dynamics in (relative) radial direction $\rho$, thus, becomes effectively one-dimensional(!). In double polar coordinates
CMS $(R, \phi)$ and relative $(\rho, \varphi)$ coordinate systems
(i) the eigenfunctions are factorizable, all factors except for $\rho$-dependent are found analytically, they have definite relative angular momentum,
(ii) dynamics in $\rho$-direction is described by a funnel-type potential and it is characterized by the hidden $sl(2)$ algebra;
(iii) at some discrete values of dimensionless magnetic fields $b\equiv \frac{B}{4\,m_r^2\,e^3\,c}  \leq 1$ the algebra $sl(2)$ emerges in finite-dimensional representation, thus, the system becomes {\it quasi-exactly-solvable}. This case has been analyzed in \cite{ET-q:2013} (see also \cite{Taut2}).

However, for the case of moving system $\mathbf {P} \neq 0$, an immediate observation is that the relative angular momentum ${\hat {\ell}}_z$ is not conserved. Consequently, the relative coordinates are not separated and the system is not reduced to one-dimensional dynamics. Neither factorization of eigenfunctions nor a hidden algebraic structure occurs. The problem is essentially two-dimensional and we arrive at the question how to solve it. A simple idea that we are going to employ is to combine a WKB expansion at large distances with perturbation theory near the minima of the potential into an interpolation. The main practical goal of this paper is to construct such an approximation for the ground state of the Hydrogen atom and then use it as variational trial function.


\section{Atom at small magnetic fields and small Pseudomomentum}
\bigskip
Let us first consider the case of weak magnetic fields and small Pseudomomentum. In this regime the interaction of particles with the magnetic field as well as magnetic field self-interaction can be treated as a perturbation of the Hydrogen atom. We focus our analysis on the ground state only. The wave function depending on relative coordinate ${\boldsymbol \rho}$ (see (\ref{psik})) can be represented in the form:
\begin{equation}
\psi\ =\ e^{-\Phi(\rho,\,\varphi)}\ .
\label{varpsi}
\end{equation}
We assume that the phase $\Phi(\rho,\,\varphi)$ has no singularities on the real plane.
Substituting (\ref{varpsi}) with CM momentum chosen to be directed along $y$-axis, thus,
${\bf P} = (P,\frac{\pi}{2})$ in polar coordinates, into the Schr\"{o}dinger equation (\ref{HKa})
we arrive at the Riccati type equation
\begin{equation}
 \frac{1}{2\,m_r}\Delta \Phi - \frac{i \,e\,B\,|\mu_2-\mu_1|}{2\,m_r}\frac{\pa \,\Phi}{\pa \varphi} - \frac{1}{2\,m_r}{(\nabla \Phi)}^2 = {\hat E}-V\ ,
\label{Riccati0}
\end{equation}
with a non-trivial potential
\begin{equation}
V \ =\  \frac{e^2\,B^2\,\rho^2}{8\,m_r}-\frac{e\,B\, P\,\rho\,\cos\varphi}{M} -\frac{e^2}{\rho}\ ,
\label{Vcos}
\end{equation}
where ${\hat E} = E-\frac{P^2}{2\,M}$.
\bigskip
Now we develop a perturbation theory for $\Phi$ and ${\hat E}$
\begin{equation}
\begin{aligned}
&\Phi \ = \ \sum_{n,k=0}^{\infty} B^n\,P^k\, \Phi_{n,k}\ ,
\\ & {\hat E}\ = \ \sum_{n,k=0}^{\infty} B^n\,P^k\, {\hat E}_{n,k} \ .
\label{PerT}
\end{aligned}
\end{equation}
Substituting (\ref{PerT}) into (\ref{Riccati0}), and collecting the terms of the order of $B^n\,P^k$, we obtain the following equation for corrections, thus, $n+k>0$:
\begin{equation}
 \frac{1}{2\,m_r}\De \Phi_{n,k} - \frac{i\ e\,B\,|\mu_2-\mu_1|}{2\,m_r}\frac{\pa}{\pa \varphi}\,\Phi_{n-1,k}
 - \frac{1}{m_r}\,\nabla \Phi_{0,0}\,\nabla \Phi_{n,k} = {\hat E}_{n,k} -Q_{n,k}\ ,\
\label{Riccati0n}
\end{equation}
where formally $\Phi_{-1,k}=0$, here $Q_{n,k}$ plays the role of an effective perturbation potential and is given by the following formulas:
\[
 Q_{0,1}\,=\,Q_{1,0}\,=\,Q_{0,2}\,=\,0\,, \quad Q_{1,1}\,=\,-\frac{e\,\rho\,\cos\varphi}{M}\,, \quad Q_{2,0}\,=\,\frac{e^2\,\rho^2}{8\,m_r}\ ,
\]
and for $n+k>2$
\begin{equation}
  Q_{n,k}\ =\ -\frac{1}{2\,m_r} \sum_{0 \leq m \leq n , 0 \leq p \leq k} \nabla \Phi_{m,p}\,\nabla \Phi_{n-m,k-p}\ ,\
\label{Qnk}
\end{equation}
where the summation is performed such a way that $m+p \neq 0$ and $m+p < n+k$. The first corrections in (\ref{PerT}) can be calculated explicitly (see Table~\ref{Zeeman2D} for some concrete cases) and the energy expansion has a form
\begin{equation}
\begin{aligned}
 {\hat E}\ =&\  -2\,m_r + \frac{3}{64\,m_r^3}\,B^2 -\frac{21 }{256\,m_r^3\,M^2}\,B^2\,P^2 -\frac{159\,{(4\,m_r+M\,\mu^2)}^2}{65536\,m_r^7\,M^2}\,B^4
   \\ &
 + \frac{3\,(115\,M\,\mu^2-2062\,m_r)}{131072\,m_r^7\,M^3}\,B^4\,P^2 + \frac{17877}{1048576\,m_r^7\,M^4}B^4\,P^4 \\ &
 -\frac{1293475\,M^2\,\mu^4+1624212\,m_r\,M\,\mu^2-1816848\,m_r^2}{805306368\,m_r^{11}\,M^4}\, B^6 \,P^2\, +\, \ldots
\label{PTenergy}
\end{aligned}
\end{equation}
\bigskip
where we put $e=1$ and $\mu=\mu_1-\mu_2$. The coefficient functions in the phase expansion (\ref{PerT}) are
\begin{equation}
\begin{aligned}
& \Phi_{0,0}=2\,m_r\,\rho   \ , \qquad  \Phi_{1,1}=-\frac{(3+4\,m_r\,\rho)\,\rho\,\sin \varphi}{16\,m_r\,M}\ , \\ & \\ &
 \Phi_{2,1}=-\frac{i\,\mu\,(51+4\,m_r\,\rho(17+8\,m_r\,\rho)\,)\,\rho\,\cos \varphi}{1536\,m_r^3\,M}
\ , \qquad  \Phi_{2,0}=\frac{\rho^2\,(4\,m_r+M\,\mu^2)(9+8\,m_r\,\rho)}{384\,m_r^2\,M}\ ,
\\ & \\ & \Phi_{0,2}=\frac{\rho^2\,(-99-40\,m_r\,\rho+3(11+8\,m_r\,\rho)\cos 2\varphi)}{3072\,m_r^2\,M^2}\ .
\label{PTphase}
\end{aligned}
\end{equation}

Since in the potential (17) the magnetic field $B$ appears as a coefficients in front of the most singular term $\sim \rho^2$ at large distances the Dyson instability phenomenon occurs when $B^2$ is negative (for discussion, see \cite{Turbiner:1979-84}). Hence, it is evident that this perturbation theory in powers of $B$ is divergent, there exists a singularity at $B=0$. Otherwise, for any fixed $B \neq 0$
the Dyson instability phenomenon does not occur when $P$ changes sign. It indicates that the series in $P$ has a finite radius of convergence, which is, of course, of the order of $B$. Thus, a domain of applicability is limited to the case
of fairly low $B$ and $P \lesssim B$.
\begin{table}[htb]
\begin{center}
\small
\begin{tabular}{|c||c|c|c|}
\hline
$\hat E\, (\text{Hartrees})\,$    &   H  ($M<\infty$)  &  H  ($M\rightarrow \infty$) & Ps $(M=2\,a.u.)$ \\ \hline
\hline
$\hat E_{0,0}$     &  $  -1.9989     $                 & $ -2  $                   & $-1$ \\
\hline
$\hat E_{2,0}$     & $ 0.04695 $                       &   $\frac{3}{64}$          & $\frac{3}{8} $\\
\hline
$\hat E_{4,0}$     &  $ -0.002435 $                    & $-\frac{159}{65536}$      & $-\frac{159}{512}$  \\
\hline
$\hat E_{2,2}$     &  $  -2.4344  \times 10^{-8}$      & $  0  $                   & $-\frac{21}{128}$ \\
\hline
$\hat E_{4,2}$     & $   0.7735 \times 10^{-9}$        & $  0  $                   & $-\frac{3093}{8192} $\\
\hline
$\hat E_{4,4}$     &   $1.502  \times10^{-15}$         & $  0  $                   & $\frac{17877}{131072}$  \\
\hline
\hline
\end{tabular}
\end{center}
\caption{Coefficients ${\hat E_{i,j}}$ in (\ref{PerT}) for the Hydrogen atom and Positronium, $M$ is total mass of the system.}
\label{Zeeman2D}
\end{table}


\section{Moving atom in a arbitrary magnetic field and Pseudomomentum}
\bigskip
In this section we consider the neutral system in a magnetic field for arbitrary values of a magnetic field and Pseudomomentum.
\bigskip
\subsection{Scaling relations}

One can relate two different neutral systems, ($e,\,m_1,\,m_2$) and
($\tilde e,\,\tilde m_1,\,\tilde m_2$). In order to do it let us make a scale transformation $\rho \rightarrow a\,\rho$ in (\ref{HKa}) and choose
\[
a\ =\ \frac{\tilde e^2\,\tilde m_r}{e^2\,m_r}   \ ,
\]
together with
\begin{equation}
\begin{aligned}
& \frac{\tilde B}{\tilde e^3\,\tilde m_r^2} \ =\ \frac{B}{e^3\,m_r^2}
 \\ &  \frac{\tilde P}{(\tilde m_1+\tilde m_2)\,\tilde e^2}\ =\ \frac{P}{(m_1+m_2)\,e^2}
 \\ &  |\tilde \mu_2 - \tilde \mu_1|\ = \  |\mu_2-\mu_1| \ . \non
\end{aligned}
\end{equation}
Then the following scaling relations emerge
\[
\psi_{{}_{\mathbf P}}(\tilde e,\,\tilde m_1,\,\tilde m_2,\,\tilde B,\,\tilde P\,;\,a\rho,\,\varphi) \ =\
\psi_{{}_{\mathbf P}}(e,\,m_1,\,m_2,\,B,\,P;\,\rho,\,\varphi) \ ,
\]
\begin{equation}
\label{sc-relqK}
 \frac{1}{\tilde e^4 \, \tilde m_r}\,
 \hat E(\tilde e,\,\tilde m_1,\,\tilde m_2,\,\tilde B,\,\tilde P) \ = \  \frac{1}{e^4 \,m_r}\,  \hat E(e,\,m_1,\,m_2,\,B,\,P)  \ .
\end{equation}

It is worth mentioning that similar scaling relations can be written for three-dimensional case \cite{Herold}.

\bigskip
\subsection{The effective potential and optimal gauge.}
\begin{figure}[htp]
\begin{center}
\includegraphics[width=4.5in, angle=0]{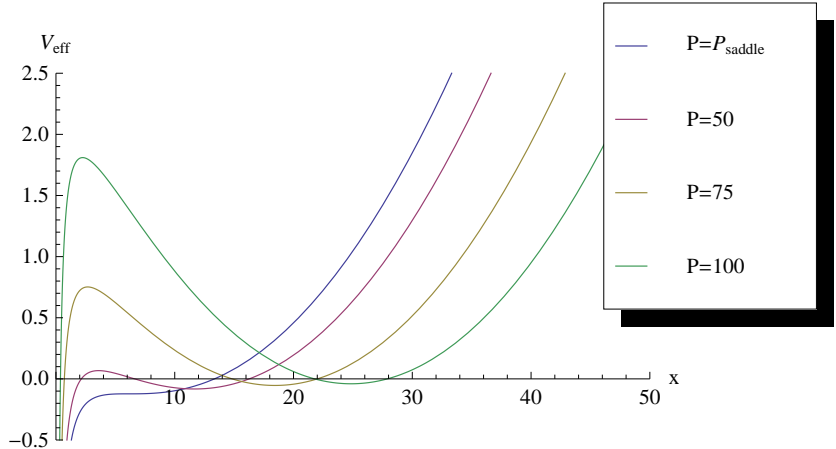}
\caption{Hydrogen atom: effective potential $V_{eff}$ (\ref{Veff}) at $P=P_{saddle},50,75,100$. $P_{saddle}=36.7$ and
$B = B_0 = 9.3918 \times 10^9 $\,G\,.}
\label{VB1}
\end{center}
\end{figure}

\bigskip
The form of the Schr\"odinger equation (\ref{HKa}) implies the existence of the gauge-invariant effective potential
\[
V_{eff}=\frac{ {( \mathbf {P}\,-\,e\,\mathbf B\times {\boldsymbol \rho} )}^2}{2\,M}   -\frac{e^2}{\rho}
\]
which describes the relative motion
\footnote{It is worth mentioning that in three-dimensional case after pseudoseparation of variables the Schr\"odinger equation describing the relative motion is also characterized by gauge-invariant effective potential \cite{CS:1993}}. In Cartesian coordinates choosing without loss of generality the CM momentum
$\mathbf P = P\,{\mathbf {\hat y}}$, we have
\bigskip
\begin{equation}
V_{eff}(x,y) \ =\ \frac{P^2}{2\,M}+\frac{e^2\,B^2}{2\,M}(x^2+y^2)-\frac{e\,B\,P}{M}x-\frac{e^2}{\sqrt{x^2+y^2}} \ .
\label{VeffTot}
\end{equation}
For any $P$ the potential $V_{eff}$ has a minimum at $x=y=0$, it corresponds to the Coulomb singularity. It can be called the Coulomb minimum. For a certain values of $P$, larger than some critical Pseudomomentum (CMS momentum), $P_{saddle}$ another minimum can occur. It is located along the line perpendicular to the direction of Pseudomomentum. In this direction, at $y=0$, $V_{eff}$ reads
\begin{equation}
  V_{eff}(x,0) \ =\ \frac{P^2}{2\,M}+\frac{e^2\,B^2}{2\,M} x^2-\frac{e\,B\,P}{M} x -
  \frac{e^2}{\mid x\mid } \ .
\label{Veff}
\end{equation}
and the position $x_0$ of minimum is given by a solution of the cubic equation
\begin{equation}
\begin{aligned}
x_0^3-\frac{P}{e\,B}\,x_0^2+\text{sign}[x_0]\frac{M}{B^2}=0\ .
\label{Vy}
\end{aligned}
\end{equation}
All three solutions of (\ref{Vy}) are real if
\begin{equation}
\begin{aligned}
  P\ \geq \  P_{saddle}\ \equiv \ {\bigg(\frac{27}{4}e^3\,B\,M\bigg)}^{\frac{1}{3}}\ .
\label{Psadd}
\end{aligned}
\end{equation}
At $P=P_{saddle}={(\frac{27}{4}e^3\,B\,M)}^{\frac{1}{3}}$ the Eq.(\ref{Vy}) has double zero which corresponds to the appearance of the saddle point in (\ref{Veff}).
It is located at $x_{saddle} = {(\frac{2\,M}{B^2})}^{\frac{1}{3}}$ (which is charge independent(!), $e>0$). For $P > P_{saddle}$, the potential (\ref{Veff}) has two minima
(see Fig.\ref{VB1}). For fixed $B$ in the limit $P \rar \infty$ we can easily obtain from (\ref{Vy}) the expression
\begin{equation}
\begin{aligned}
       x_{0,min} \approx \frac{P}{e\,B}- \frac{e^2\,M}{P^2} + \ldots \ ,
\label{min}
\end{aligned}
\end{equation}
therefore, $x_{0,min}$ grows linearly at large $P$ and $V_{eff}(x_{0,min},0)$ tends to zero as $-\frac{B\,e^3}{P}$. Similarly, the position of the maximum  of the effective potential $V_{eff}(x,0)$,
\begin{equation}
\begin{aligned}
  x_{0, max}\approx\sqrt{\frac{e\,M}{P\,B}} + \frac{e^2\,M}{2\,P^2}\ +\ \ldots
\label{max}\ ,
\end{aligned}
\end{equation}
thus, $x_{0,max}\rar 0$ with growth of $P$ and $V_{eff}(x_{0,max},0) \rar \infty$\,.
The behavior of the barrier height $\De V=V_{eff}(x_{0,max},0)-V_{eff}(x_{0,min},0)$ at large $P$
is given by the expansion
\begin{equation}
\begin{aligned}
 \De V \ = \ \frac{P^2}{2\,M} - \sqrt{\frac{4\,B\,e^3\,P}{M}} + \frac{3\,B\,e^3}{2\,P}  +
 \sqrt{\frac{B^3\,e^9\,M}{16\,P^5}}    + \ldots
\label{DV}
\end{aligned}
\end{equation}

For $P=0$ the second minimum in $V_{eff}$ (\ref{Veff}) does not exist and the potential possesses azimuthal symmetry. In this case, the symmetric gauge emerges naturally as the most convenient. The
convenience is related with the fact that for this gauge the ground state eigenfunction is real. For $P\neq 0$ the azimuthal symmetry is broken, consequently, the most convenient choice of the gauge to treat the problem is no longer evident. A question can be posed: in what gauge the ground state eigenfunction is real? In such a gauge the trial function for the ground state can be searched among real functions. This strategy was realized in \cite{PRepts}.

Since we are going to use an approximate method for solving the Schr\"{o}dinger equation with the Hamiltonian (\ref{HKa}), a quality of the approximation of ground state function can depend on the gauge. In particular, one can ask whether can one find a gauge for which a given trial function leads to minimal variational energy. Such a gauge (if found) can be called optimal for a chosen trial function.

To this end, it is convenient to introduce a unitary transformation
\begin{equation}
\label{UU}
  {U}\ =\ e^{-i\,{e\,(\mu_2-\mu_1)}\,\mathbf A_{{\boldsymbol \rho}_{0}} \cdot {\boldsymbol \rho}}\ ,
\end{equation}
where
\begin{equation}
\label{cengaug}
{\boldsymbol \rho}_{0}=\frac{d\,P}{e\,B}\,{\mathbf {\hat x}}
\end{equation}
and $0 \leq d \leq 1$ is a parameter. The unitary transformed Hamiltonian
(\ref{HKa}) takes the form
\begin{equation}
  {h}_{d} \ \equiv\  U^{-1}\,{h}\, U \ = \ \frac{{({\mathbf {\hat p}}-e\,(\mu_2-\mu_1)\,
  {\mathbf A_{(\boldsymbol \rho -{\boldsymbol \rho}_{0}   )}})}^2}{2\,m_{r}} +
  \frac{(\mathbf {P}+e\,{\mathbf B \times {\boldsymbol \rho}}  )^2}{2\,M} -
  \frac{e^2}{\rho}  \ ,
\label{HP}
\end{equation}
This transformation implies that we consider now the Schr\"odinger equation in a linear gauge for which the position of the gauge center, where $\mathbf A(x,y)=0$, is located at $y=0,\,
x=\frac{d \,P}{e\,B}$. For $P > P_{saddle}$ we expect the gauge center to be localized on the line $y=0$, between the origin $x=0$ and the second minimum $x=x_{0,min}$ of $V_{eff}$, (see (\ref{min})). Thus, the vector potential can be considered as a variational function and can be chosen by a procedure of minimization as it was proposed in \cite{PRepts} (see also for discussion \cite{Vincke}). The case $d = 0$ has been used in the past to study the so-called centered states with wavefunction peaked at the Coulomb minimum \cite{Burkova}. While for the so-called "decentered" states it seems natural to consider $d = 1$. The eigenvalue problem
\begin{equation}
{h}_{d}\,\chi_{{}_{\mathbf P}}= E \,\chi_{{}_{\mathbf P}} \ ,
\label{hxi}
\end{equation}
where $\chi_{{}_{\mathbf P}} = e^{i\,{e\,(\mu_2-\mu_1)}\,\mathbf A_{{\boldsymbol \rho}_{0}} \cdot {\boldsymbol \rho}} \,\psi_{{}_{\mathbf P}} $, is the central object of our study hereafter.

\bigskip

\subsection{Asymptotics.}

\vspace{0.2cm}

If we put $\chi_{\mathbf P} = \text{e}^{-\varphi}$ in (\ref{hxi}),
one can construct the WKB-expansion at large $\rho = \sqrt{x^2+y^2}$ for the phase $\varphi$.
The leading term at $\rho \rar \infty$ is given by
\begin{equation}
     \varphi \ =\  \frac{e\,B}{4}\,\rho^2  +  O(\rho) \ .
\label{large}
\end{equation}
Assuming the condition (\ref{Psadd}) is fulfilled, the potential (\ref{Vy}) has the second minimum at $x_0 \neq 0$. It can be constructed the double Taylor expansion of the phase at $x = x_{0}, y=0$\ ,
\begin{equation}
\varphi \ =\  \al_0+\al_2\,{(x-x_{0})}^2  + \al_3\,{(x-x_{0})}^3 + \al_4\,{(x-x_{0})}^4 +  \ldots   + \beta_2\,y^2+\beta_4\,y^4 + \ldots + \gamma_3\,(x-x_{0})\,y^2 +   \ldots
\label{x0}
\end{equation}
where $\al_0$ is unimportant constant and
\[ \al_2  = E_x\,m_r\ ,\ \al_3  =  \frac{e^2}{6\,E_x\,x_0^4}\ ,\
\al_4 = \frac{1}{6}\,m_r\ (2\,E_x^2\,m_r-\frac{e^2\,B^2}{2\,M}+\frac{e^2}{x_0^3})\ ,
\]
\[    \beta_2  = E_y\,m_r\ ,\
\beta_4 = \frac{1}{6}\,m_r\,(2\,E_y^2\,m_r-\frac{e^2\,B^2}{2\,M}-\frac{e^2}{2\,x_0^3})\ , \
\gamma_3 = - \frac{e^2}{2\,E_x\,x_0^4}\ ,
\]
here $E =E_x+E_y$.
Similarly, at $x=0, y=0$,
\begin{equation}
\varphi \ =\   2\,m_r\,\rho + O(\rho^2)\ .
\label{origin}
\end{equation}


\subsection{Approximations}

Making interpolation between WKB-expansion (\ref{large}) and the perturbative expansion (\ref{origin}), (\ref{x0}), respectively, we arrive at the following expressions
\begin{equation}
\begin{aligned}
 &    \vphi_c = \ \frac{A_0 + A_1\,{\rho} + A_2\,{x}\,{\rho} +  A_3^2\,{\rho^3}
}{   \sqrt{ 1 + A_4\,{x} + A_5^2 \,{\rho^2}}   } - \frac{\alpha_c}{2}\,\log(1+A_4\,{x} + A_5^2\,{\rho^2})
\\  &  \\ & \vphi_m = \ \frac{D_0 + D_1\,{{\tilde x}^2}+D_2\,{y^2} +D_3^2\,{\varrho^4}}{
  \sqrt{1 + D_4\,{{\tilde x}^2} +  D_5\,{y^2}  +  D_{6}^2\,{\varrho^4}}}
\label{phis}
\end{aligned}
\end{equation}
where $\tilde x = x-x_m$, $\varrho^2 = {\tilde x}^2+y^2$ and $A^\prime s,\, D^\prime s\,, \alpha_c\,,\,x_m$ are non-linear variational parameters.
Following the prescription formulated in \cite{Turbiner:1988-2010} we construct a trial function for
the ground state in a form of linear superposition
\begin{equation}
\chi_{\mathbf P}  \ = \ C_1 \,e^{-\vphi_c} + C_2\,e^{-\vphi_m}\ ,
\label{interp}
\end{equation}
where $C^\prime s\,$ are linear variational parameters. Supposedly, they should behave smoothly as a function of a magnetic field.



\section{Results}

\hskip 1cm
We carried out a variational study of the two-body neutral system on a plane moving across a magnetic field. The main emphasis is to explore stability of the moving system, thus, studying the ground state. For the case of Hydrogen atom, the energy for several magnetic fields $0 < B < 1000$\,a.u. and values of Pseudomomentum (= CMS momentum $P$) $0 \leq P < 200$ a.u. is presented in Table \ref{Table1}. The energy grows monotonically and rather sharp as a function of a magnetic field for fixed Pseudomomentum but at much slow pace as a function of Pseudomomentum for a fixed magnetic field. It is worth noting that for fixed $B$ the energy $E$ as a function of $P$ tends asymptotically to the ground state energy of two non-interacting charges in a magnetic field.

\begin{figure}[htp]
\centering
\subfigure[]{\includegraphics[width=3.5in,angle=-90]{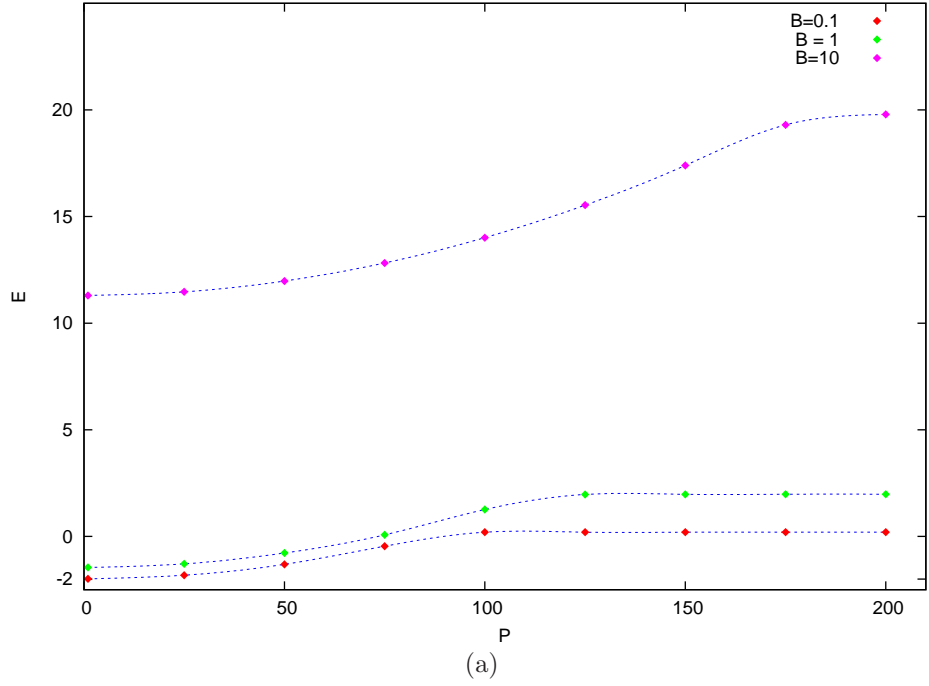}} \qquad \subfigure[]{\includegraphics[width=3.5in,angle=-90]{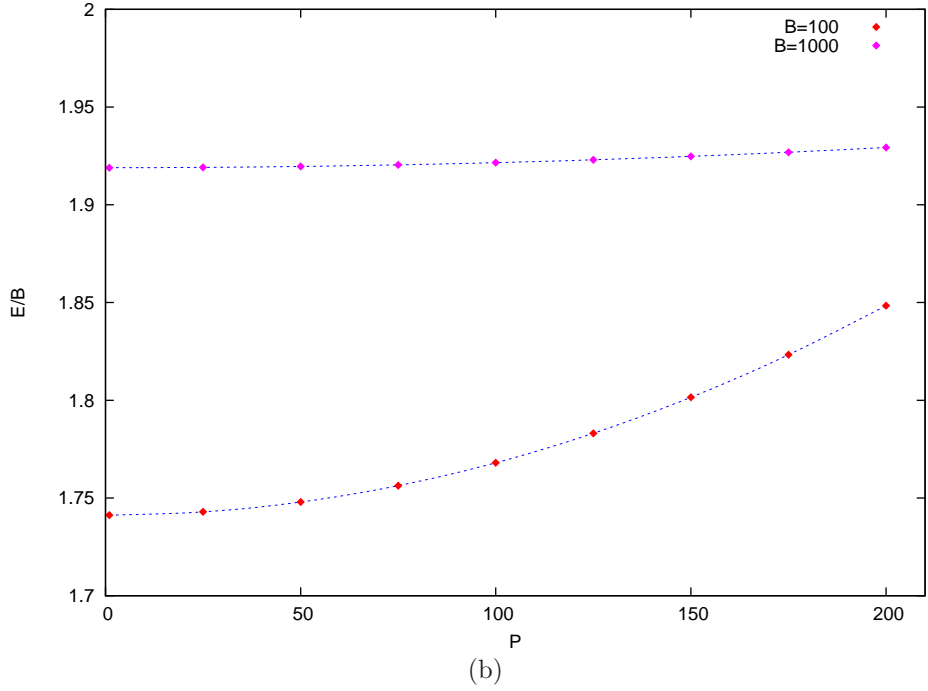} }
\caption{The energy (in Hartrees) of ground state {\it vs} Pseudomomentum $P$. (a) $E$ \textit{vs} $P$ for magnetic fields $B=0.1,\,1,\,10$. (b) $E/B$ \textit{vs} $P$ for magnetic fields $B=10^2,\,10^3$\,a.u.; magnetic field $B$ and Pseudomomentum $P$ in effective atomic units, $B_0 = 9.3917\times 10^9$\,G , $P_0/\hbar = 3.47 \times 10^{11}\, { \rm cm}^{-1}$, respectively.
 }
\label{E0}
\end{figure}

\begin{figure}[htp]
\begin{center}
\includegraphics[width=4.0in,angle=-90]{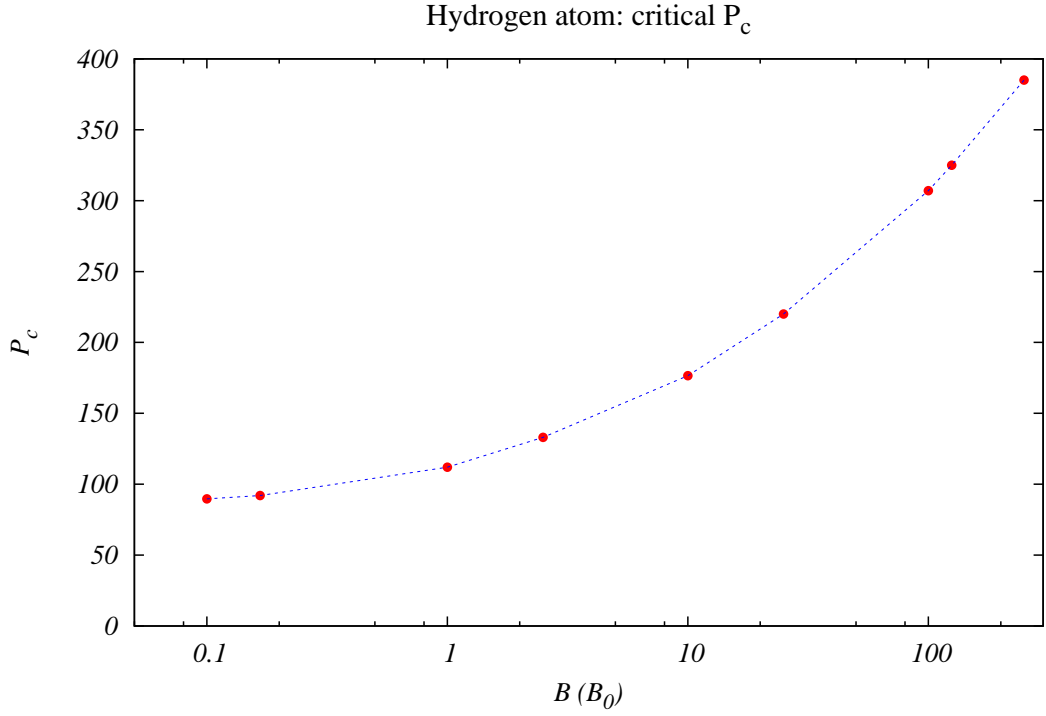}
\caption{Critical Pseudomomentum $P_c$ \textit{vs} $B$\,. }
\label{Pcr:fig.3}
\end{center}
\end{figure}

\begin{figure}[htp]
\begin{center}
\includegraphics[width=3.0in,angle=0]{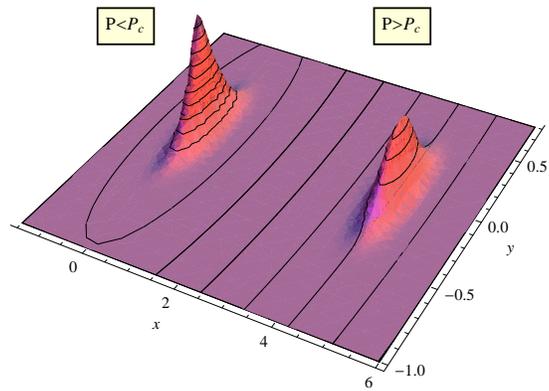}
\caption{Hydrogen atom: the variational density of the ground state function $|\chi_{{}_{\mathbf P}}|^2$ of Hamiltonian (\ref{hxi}) at $P < P_{c}$ (left peak) and $P > P_{c}$ (right peak) for $B=10$\,a.u. Magnetic field in effective atomic units, $B_0 = 9.3917 \times 10^9 \,G$}
\label{Pc:fig.4}
\end{center}
\end{figure}

\hskip 1cm
After making a minimization, one can see the appearance of a sharp change in behavior of parameters in (\ref{interp}) as a function of $P$. It is related with a fact of the existence of a certain
critical Pseudomomentum $P_c > P_{saddle}$ such that for $P < P_c$ the optimal linear parameters
$C_1 \approx 1,\,C_2 \approx 0$ the wavefunction has a peak near the Coulomb singularity (centered state). At $P > P_c$ the situation gets opposite: the parameters $C_1 \approx 0,\,C_2 \approx 1$ and the wavefunction is peaked near the second well of (\ref{Veff}), see Fig.\ref{VB1} (we call this well the {\it magnetic} well) which corresponds to a decentered state. The existence of such a change in the behavior of parameters results also in a specific behavior of energy dependence (see for example Fig.\ref{E0}) and mean interparticle separation vs the Pseudomomentum. From physical point of view at $P = P_c$ the effective depth of the Coulomb well and one of the magnetic well get equal. If $P < P_c$ the effective depth of the Coulomb well is larger (or much larger depending on a magnetic field strength) than one of the magnetic well, the system prefers to stay at the Coulomb well. If $P > P_c$ the effective depth of the Coulomb well is smaller (or much smaller depending on a magnetic field strength) than one of the magnetic well, the system prefers to stay at the magnetic well. For all studied magnetic fields the barrier between wells is very large, the probability of tunneling from one well to another one is very small. Hence, the energy behavior {\it vs.} CMS momentum is defined by the behavior in one well or another, it is close to classical behavior. Thus, the presence of the second minimum in the effective potential can be neglected. This phenomenon is very well-pronounced
at small magnetic fields. Similar behavior was observed in three-dimensional case in \cite{CS:1993} (for Hydrogen atom) and \cite{Ackermann:1997} (for Positronium).
The behavior of $P_c$ as a function of magnetic field is presented in Fig. \ref{Pcr:fig.3}, $P_c$ grows with an increase of $B$.
Typical distribution $|\chi_{{}_{\mathbf P}}|^2$ for $P < P_c$ and $P > P_c$ is shown on Fig. \ref{Pc:fig.4}.

\hskip 1cm
The evolution of the gauge center parameter $d$ (see (\ref{cengaug})) {\it vs.} CMS momentum is shown in Table \ref{Table2}. For $P < P_c$ the parameter $d$ is very small, it varies within $[0 - 0.051]$ for magnetic field range $0.1 - 1000$\,a.u. while for $P > P_c$ it is close to 1, it varies within $[0.985 - 1]$ for magnetic field range $0.1-10$\,a.u. Thus, the gauge changes from the symmetric gauge centered at the Coulomb well to symmetric gauge but centered at the magnetic well. In turn, the parameter $d$ remains almost equal to $0$ up to $P = P_c$ (which means the gauge center coincides with a position of the Coulomb singularity, then sharply jumps to a value close to $1$ (gauge center coincides with a position of the minimum of magnetic well), displaying a behavior which looks like a phase transition. But it is not a phase transition: the energy changes sharply but smooth. For $P = P_c$ the parameter $d \sim 0.5$. There exists a certain domain of transition from one regime to another. Overall situation looks very similar to one for the $H_2^+$ molecular ion in a magnetic field in inclined configuration \cite{PRepts}.

\hskip 1cm
In order to illustrate the transition from a centered state to a decentered one we have calculated, using the trial function (\ref{phis}) with optimal parameters, the expectation value of the relative coordinate, $\langle \rho \rangle$, see Table \ref{exprho}. At weak magnetic fields $B$, the transition is very sharp, becoming even more pronounced with a magnetic field decrease. For all studied magnetic fields and fixed CMS momentum $\langle \rho \rangle$ is finite. Furthermore, the trial function (\ref{phis}) remains normalizable in relative coordinates ${\boldsymbol \rho}$, see (\ref{psik}). It indicates stability of the neutral system.

\hskip 1cm
To complete the study we show in Tables VII - XIII (see Supplementary Materials) the nonlinear parameters $A$'s, $D$'s and $\al_c$ of the trial function (\ref{phis}) as a function of the magnetic field strength for the optimal configuration. For all considered values of Pseudomomentum the parameter $A_1=1.998,\,1.998,\,1.982,\,1.965,\,1.938$ at $B=0.1,1,10,100,1000$\,a.u., respectively. A deviation $|A_1-2|$ measures (anti)-screening of the electric charge due to the presence of a magnetic field. The optimal value of energy corresponds to $x_m \approx x_{0,min}$.

\hskip 1cm
Our variational results are checked on agreement with results obtained in other methods.
Comparison with perturbative theory in the high-magnetic-field limit (taking the Coulomb
interaction as a perturbation) is presented in Table \ref{Eper}. Let us note that with the increase of a magnetic field the variational and perturbative results get closer and for $B=1000$\,a.u.
they agree in four significant digits.
We adopted the Lagrange mesh method based on the Vincke-Baye approach \cite{Baye} to obtain the ground energy for $B=1$\,a.u. and different Pseudomomenta $P$, see Table \ref{TabMesh}. For all studied values of $P$ the variational energy is in agreement with Lagrange mesh calculations in not less than 5 s.d.

\begin{figure}[htp]
\begin{center}
\includegraphics[width=5.0in,angle=-90]{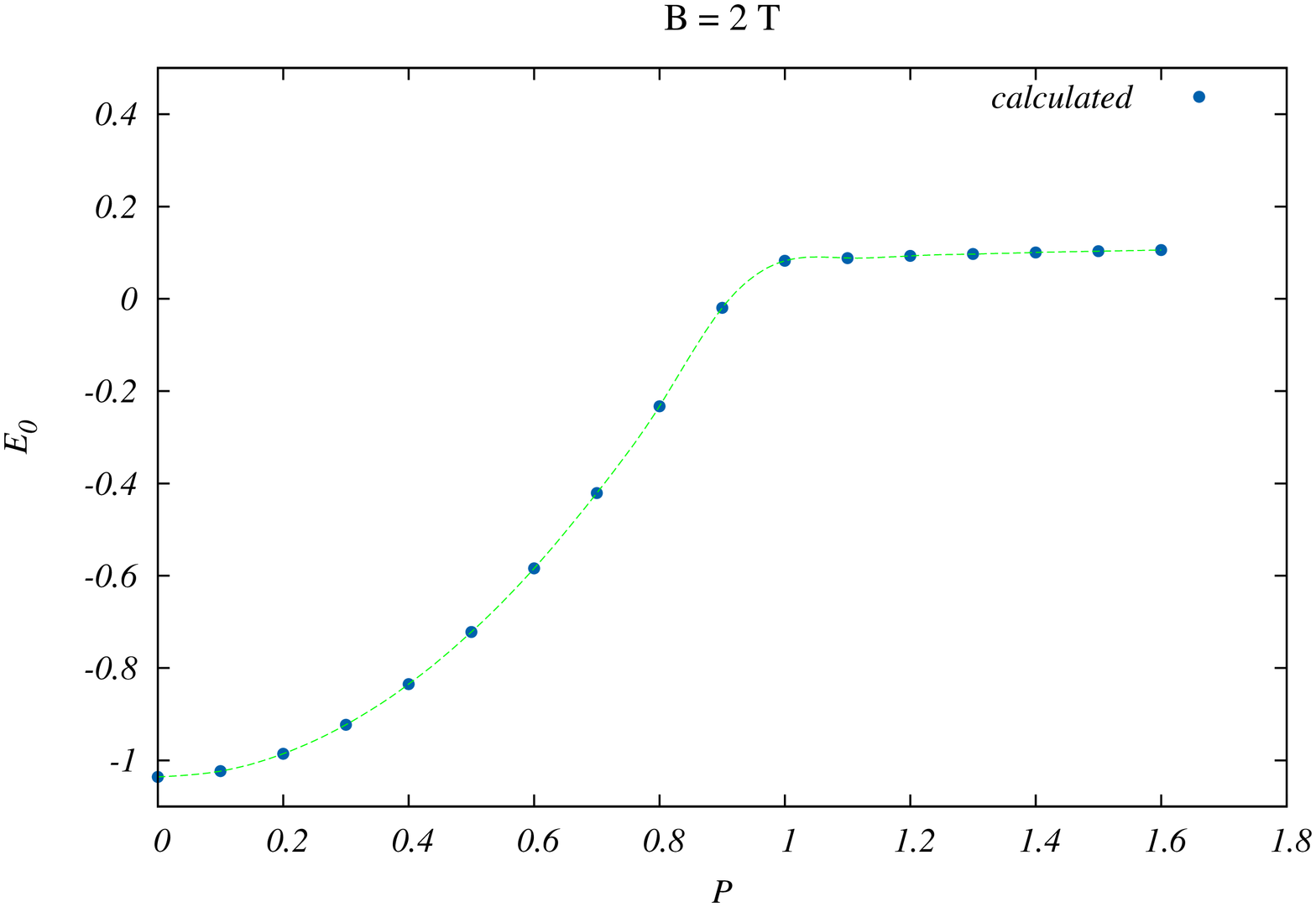}
\caption{Ground state energy: Direct excitons in GaAs$/$A$\text{l}_{0.33}$G$\text{a}_{0.67}$As~CQW\'s with $\epsilon=12.1$ at $B=2\,T$ studied in \cite{Lozovik:2002}, see Fig.5. Present results are marked by bullets. The units used are the same as in \cite{Lozovik:2002}. }
\label{compar:fig.4}
\end{center}
\end{figure}

It is quite interesting to make a comparison with a single numerical study by Lozovik \textit{et al.} \cite{Lozovik:2002} of direct excitons in GaAs$/$A$\text{l}_{0.33}$G$\text{a}_{0.67}$As CQW's.
This study was done for very small values of magnetic field only.
It can be seen that variational energies based on (\ref{phis}) are systematically lower in $\sim 10\%$. Qualitative behavior of variational energy as a function of Pseudomomentum $P$ is similar to one presented in \cite{Lozovik:2002}, see Fig. \ref{compar:fig.4}.


\begin{center}
\end{center}
\setlength{\tabcolsep}{11.0pt}
\setlength{\extrarowheight}{1.0pt}
\begin{table}[th]
\begin{center}
\begin{tabular}{|c||c|c|c|c|c|}
\hline
\\[-23pt]
$P$   & \multicolumn{5}{c|}{ $E$ } \\
\hline               & \multicolumn{5}{l|}{\hspace{0.0cm} $B=0.1$ \hspace{0.6cm} $B = 1$ \hspace{0.6cm}
$B=10$ \hspace{0.6cm} $B =100$ \hspace{0.6cm} $B=1000$ }\\
\hline
\hline
$0$                  &   $-1.9914$    &   $-1.4587$         &   $11.299$   &   $174.124$     &   $1918.98$
\\[3pt]
\hline
$1$                  &   $-1.9912$    &   $-1.4585$         &   $11.299$   &   $174.124$     &   $1918.98$
\\[3pt]
\hline
$25$                 &   $-1.8213$    &   $-1.2887$         &   $11.468$   &   $174.291$     &   $1919.15$
\\[3pt]
\hline
$50$                 &   $-1.3110$    &   $-0.7787$         &   $11.976$   &   $174.794$     &   $1919.63$
\\[3pt]
\hline
$75$                 &   $-0.4604$    &   $0.0713$          &   $12.823$   &   $175.631$     &   $1920.44$
\\[3pt]
\hline
$100$                &   $0.19589^d$  &   $1.261$           &   $14.009$   &   $176.803$     &   $1921.57$
\\[3pt]
\hline
$125$                &   $0.19669^d$  &   $1.966^d$         &   $15.534$   &   $178.310$     &   $1923.02$
\\[3pt]
\hline
$150$                &   $0.19722^d$  &   $1.972^d$         &   $17.398$   &   $180.152$     &   $1924.80$
\\[3pt]
\hline
$175$                &   $0.19760^d$  &   $1.976^d$         &   $19.600$   &   $182.329$     &   $1926.90$
\\[3pt]
\hline
$200$                &   $0.19789^d$  &   $1.978^d$         &   $19.788^d$ &   $184.840$     &   $1929.32$
\\[3pt]
\hline
\end{tabular}
\caption{\small Ground state energy $E$ in Hartrees (see (\ref{hxi})); magnetic field $B$ and Pseudomomentum $P$ in effective atomic units, $B_0 = 9.3917\times 10^9$\,G ,
$P_0/\hbar = 3.47 \times 10^{11}\, { \rm cm}^{-1}$, respectively. Energies corresponding to decentered states marked by ${}^d$\,(see text).}
\label{Table1}
\end{center}
\end{table}


\begin{center}
\end{center}
\setlength{\tabcolsep}{25.0pt}
\setlength{\extrarowheight}{1.0pt}
\begin{table}[th]
\begin{center}
\begin{tabular}{|c||c|c|c|c|c|}
\hline
\\[-23pt]
$P$   & \multicolumn{5}{c|}{ $d$ } \\
\hline                 & \multicolumn{5}{l|}{\hspace{0.0cm} $B=0.1$ \hspace{0.8cm} $B = 1$ \hspace{0.8cm}
$B=10$ \hspace{0.8cm} $B =100$ \hspace{0.8cm} $B=1000$ }\\
\hline
\hline
$P<P_c$                     &   $0$       &   $0.0005$    &   $0.004$   &   $0.0156$     &   $0.051$
\\[3pt]
\hline
$P>P_c$                     &   $1$       &   $1$         &   $0.985$   &   $-$          &   $-$
\\[3pt]
\hline
\end{tabular}
\caption{\small Optimal gauge parameter $d$ (see (\ref{cengaug})) for different magnetic fields and Pseudomomenta; magnetic field in effective atomic units, $B_0 = 9.3917\times 10^9$\,G.}
\label{Table2}
\end{center}
\end{table}

\begin{center}
\end{center}
\setlength{\tabcolsep}{26.0pt}
\setlength{\extrarowheight}{1.0pt}
\begin{table}[th]
\begin{center}
\begin{tabular}{|c||c|c|c|}
\hline
\\[-23pt]
$P$   & \multicolumn{3}{c|}{ $\langle \rho \rangle$ } \\
\hline                 & \multicolumn{3}{l|}{\hspace{0.0cm} $B=0.1$ \hspace{0.8cm} $B = 1$ \hspace{0.8cm}
$B=10$  }\\
\hline
\hline
$0$                     &   $0.49$       &   $0.39$        &   $0.175$
\\[3pt]
\hline
$75$                    &   $0.49$       &   $0.39$        &   $0.175$
\\[3pt]
\hline
$100$                   &   $250^d$      &   $0.39$         &   $0.175$
\\[3pt]
\hline
$125$                   &   $312^d$      &   $31^d$         &   $0.175$
\\[3pt]
\hline
$150$                   &   $375^d$     &   $37^d$         &   $0.175$
\\[3pt]
\hline
$175$                   &   $437^d$     &   $43^d$         &   $0.175$
\\[3pt]
\hline
$200$                   &   $500^d$     &   $50^d$         &   $4.9^d$
\\[3pt]
\hline
\end{tabular}
\caption{\small Expectation value $\langle \rho \rangle$: magnetic field in effective atomic units, $B_0 = 9.3917\times 10^9 \,G$. At $B=100, 1000$\,a.u.,\ $\langle \rho \rangle \approx 0.06, 0.019$, respectively. Data corresponding to decentered states marked by ${}^d$ \,(see text).
For $P<P_c$, $\langle \rho \rangle \approx 0 $, while for $P>P_c$,
$\langle \rho \rangle \approx \langle x\rangle \approx x_{0,min}\propto P$\ .}
\label{exprho}
\end{center}
\end{table}

\setlength{\tabcolsep}{25.0pt}
\setlength{\extrarowheight}{0.0pt}
\begin{table}[th]
\begin{center}
\begin{tabular}{|c||c|c|}
\hline
\\[-21pt]
$B$   & \multicolumn{2}{c|}{ $E$ } \\
\hline                 & \multicolumn{2}{c|}{$E_0$ \hspace{1.9cm} $E^{L}_{0}$ }\\
\hline
\hline
$0.1$                  &   $-1.9914$        &   $-0.592$
\\[3pt]
\hline
$1$                    &   $-1.4587$        &   $-0.506$
\\[3pt]
\hline
$10$                   &   $11.299$         &   $12.066$
\\[3pt]
\hline
$100$                  &   $174.124$        &   $174.83$
\\[3pt]
\hline
$1000$                 &   $1918.98$        &   $1919.68$
\\[3pt]
\hline
\end{tabular}
\caption{\small Ground state energy in Hartrees. $E_0$ from (\ref{phis}) and $E_0^L$ from perturbation theory. Magnetic field in effective atomic units, $B_0 = 9.3917\times 10^9$ \,G.}
\label{Eper}
\end{center}
\end{table}


\setlength{\tabcolsep}{25.0pt}
\setlength{\extrarowheight}{0.0pt}
\begin{table}[th]
\begin{center}
\begin{tabular}{|c||c|c|}
\hline
\\[-21pt]
$P$   & \multicolumn{2}{c|}{ $E$ } \\
\hline                 & \multicolumn{2}{c|}{$E_0$ \hspace{1.9cm} $E^{mesh}_{0}$ }\\
\hline
\hline
$0$                     &   $-1.45879$            &   $-1.45879$
\\[3pt]
\hline
$1$                     &   $-1.45852$            &   $-1.45852$
\\[3pt]
\hline
$25$                    &   $-1.28877$            &   $-1.28877$
\\[3pt]
\hline
$50$                    &   $-0.77872$            &   $-0.77871$
\\[3pt]
\hline
$75$                    &   $0.07136$             &   $0.07137$
\\[3pt]
\hline
$100$                   &   $1.26149$             &   $1.26149$
\\[3pt]
\hline
\end{tabular}
\caption{\small Ground state energy at $B=B_0$. $E_0$ from (\ref{phis}) and $E_0^{mesh}$ obtained with the Lagrange mesh method (parameter $d$ taken from the corresponding variational results based on (\ref{phis})).
Magnetic field in effective atomic units, $B_0 = 9.3917\times 10^9 \,G$.}
\label{TabMesh}
\end{center}
\end{table}

\section{Conclusions}

\hskip 1cm
Summarizing, we want to state that a simple uniform approximation of the ground state eigenfunction for the two-dimensional moving Hydrogen atom and Positronium in a magnetic field is constructed.  It manifests an approximate solution of the problem. The key element of the procedure is to make an interpolation between the WKB expansion at large distances and perturbation series at small distances both for the phase of the wavefunction; in other words, to find an approximate solution for the corresponding eikonal equation. In general, the separation of variables helps us to solve this problem easily. In our case of non-separability of variables the WKB expansion of a solution of the eikonal equation cannot be constructed in a unified way, since it depends on the way to approach to infinity. However, a reasonable approximation of the first dominant growing terms of the WKB expansion of the phase of wavefunction seems sufficient to construct the interpolation between large and small distances giving the results of rather high accuracy.

\hskip 1cm
It was demonstrated that for all magnetic fields and all CMS momenta a neutral system which moves across a magnetic field along the plane is bound. Its energy grows with a magnetic field strength increase as well as CMS momentum increase. For fixed magnetic field $B$ energy behavior demonstrates a sharp change for a certain value of CMS momentum $P_c (B)$. It seems this phenomenon can be used to measure a magnetic field strength. This effect was already mentioned in the case of three-dimensional Hydrogen atom moving across magnetic field \cite{Potekhin}.

\hskip 1cm
We calculated the corrections to the energy and to the wavefunction for the ground state of the Hydrogen atom and Positronium in a double perturbation theory in $B$ and $P$  since the ground state remains nodeless. This became possible as a result of the application of the {\it nonlinearization} method, since it is difficult to perform such calculations within the framework of the standard approach with use of sums over intermediate states or Green functions. In particular, we succeeded in computing the crossterms $\sim B^n P^k$ for arbitrary masses $m_1,m_2$. It is quite non-trivial task due to non-separability of variables in original Hamiltonian. To the authors knowledge, the corrections to the energy and to the wave function of orders $B^2\,P^2$, and $B^4\,P^2$ (cross terms) are reported for the first time. Concrete results for the Hydrogen atom and Positronium are presented in Table \ref{Zeeman2D}. In general, the correction $\Phi_{n,k}$ to the phase of wave function for even $k$ is non-vanishing, while for odd $k$ it is pure imaginary(!), except for the case of equal masses, e.g. Positronium. Since the procedure exploited was purely algebraic, one can use symbolic calculation routines like MAPLE to find higher order corrections analytically.
However, it should be kept in mind that perturbation theory in $B$ and $P$ series is divergent.

\begin{acknowledgments}
  The authors are grateful to J. C. L\'opez Vieyra and H. Olivares Pilon for their interest in the present work, helpful discussions and important assistance with computer calculations.
  This work was supported in part by the University Program FENOMEC, and by the PAPIIT
  grant {\bf IN109512} and CONACyT grant {\bf 166189}~(Mexico).
\end{acknowledgments}

\end{document}